\documentclass[aps,preprint]{revtex4}%
\usepackage{amsfonts}
\usepackage{amsmath}
\usepackage{amssymb}
\usepackage{graphicx}%
\newtheorem{theorem}{Theorem}
\newtheorem{acknowledgement}[theorem]{Acknowledgement}

\begin{document}
\title{Coherence as ultrashort pulse train generator}
\author{Gevorg Muradyan and Mariam Hovhannisyan}
\affiliation{Department of Physics, Yerevan State University, 1 Alex Manookian Yerevan,
375025 Armenia}
\email{gmurad@ysu.am}


\begin{abstract}
Intense, well-controlled regular light pulse trains start to play a crucial
role in many fields of physics. We theoretically demonstrate a very simple and
robust technique for generating such periodic ultrashort pulses from a
continuous probe wave which propagates in a dispersive thermal gas media.

\end{abstract}
\maketitle

\section{\bigskip Introduction}

The invention of the optical frequency comb has revolutionized optical
frequency metrology [1-4]. Today it is playing an important role in high
resolution spectroscopy [5],the spectral purity and large bandwidth of optical
frequency combs provides also means for the precise control of generic quantum
systems such as laser cooling of molecules or exotic atomic species [6,7], and
quantum state engineering in molecules [8-10]. \ Optical frequency comb is
becoming a crucial component in the field of quantum information science,
where complex multilevel quantum systems must be controlled with great
precision [10,11] and frequency comb technique promises to become an effective
tool in astronomical observations [12].

The usage of quantum interference effects in order to manipulate the optical
properties of gaseous atomic or molecular medium has by now been established
as a useful and powerful method. In particular in [13] Harris and co-workers
have suggested and used a Raman-type three level interaction scheme in
D2-molecular gas to get series of femtosecond pulses. In our recent paper
[14], discussing propagation of radiation probe wave in a medium of dressed
two-level atoms initially prepared in a quantum superpositional state of
ground and excited energy levels, we showed that it splits into a sequence of
ultrashort pulses with easy and precise tunning of possible control
parameters. Later \ we will refer to this scheme as QS (quantum superposition)
generator. \ In general this process can be accompanied by pulse
amplification. It may be interesting, in addition, that the gas refractive
index in comparison with earlier known results contains out of dipole
approximation terms of resonant nature which have no saturation in dependence
on pump wave intensity [15].

In this Paper, we bring the QS generator problem discussion \ closer to real
experimental settings. As a crucial point on this path we see the manner of
(superposition) state preparation. \ The most convenient way to embody the
superposition, is rapid switching \ on of the dressing field. \ It does not
require additional \ perturbing sources in the experimental setup and gives
\ number of parameters (such as the switching time, pump wave intensity and
resonance detuning) to regulate the superposition. \ We present \ here the
whole chain, starting from coherent state preparation and finishing with
incident wave modulation. We show that under appropriate conditions,
spontaneous emission and Doppler broadening have small impact on the comb
generation process.

\section{The Model}

So we consider a gas of two-level atoms with energy difference $\hbar
\omega_{0}$ between the excited and ground internal atomic bare states
$\left\vert 2\right\rangle $ and $\left\vert 1\right\rangle $ in a far
off-resonance field of the pump field
\begin{equation}
\text{ }E_{pump}(z,t)=\text{ }\varepsilon_{pump}(z,t)\exp[ik_{pump}%
z-i\omega_{pump}t]+c.c.\text{,} \tag{1}\label{1}%
\end{equation}
where $\omega_{pump}$ is the carrying frequency of the pump field,
$k_{pump}=\omega_{pump}/c$ and $\varepsilon_{pump}(z,t)$ is the slowly varying
field amplitude. The spin of relevant to optical transition electron and the
possible sublevel structures are not taken into account. \ For the pump field
amplitude we will take a functional form%

\begin{equation}
\varepsilon_{pump}(z,t)=\frac{\varepsilon_{0}}{1+e^{-(t-z/c)/T}}\text{,}
\tag{2}\label{2}%
\end{equation}
which means that the pump field has a switching front characteristic duration
$T$ and travels from left to right \ along the $z$ axis. The interaction
Hamiltonian $V$ in dipole and rotating wave approximations will reproduce the
functional form given in (2). \ This form is close to real experimental pulse
turn-on process\ and what is not less important the atom-field interaction
problem has an analytical solution for it [16]. The atomic wavefunction is
given by%
\begin{equation}
\left\vert \Psi(z,t)\right\rangle _{pump}=f(t-z/c)\left\vert 1\right\rangle
e^{-\frac{i}{\hbar}E_{1}t}+g(t-z/c)\left\vert 2\right\rangle e^{-\frac
{i}{\hbar}E_{2}t} \tag{3}\label{3}%
\end{equation}
with%
\begin{equation}
f(t-z/c)=(1-u)^{\sigma}F(a,b;c;u)\text{,} \tag{4}\label{4}%
\end{equation}%
\begin{equation}
g(t-z/c)=e^{i(kz+\Delta t)}((1-u)^{\sigma}F(a,b,c;u)-(1-u)^{\sigma
+1}F(a+1,b+1;c+1;u))\text{,} \tag{5}\label{5}%
\end{equation}
where $E_{1}$ and $E_{2}$ are energies of corresponding bare states,
$\Delta=\omega_{pump}-\omega_{0}$, $\sigma=-iV_{0}T$, $u=-e^{-t/T}$,
$a=(iT/2)\left(  -\Delta-2V_{0}+\sqrt{\Delta^{2}+4V_{0}^{2}}\right)  $,
$b=(iT/2)\left(  -\Delta-2V_{0}-\sqrt{\Delta^{2}+4V_{0}^{2}}\right)  $ and
$c=-i\Delta T$ .

Propagation of a weak probe field through a medium of atoms can be described
by wave equation%
\begin{equation}
\left(  \nabla^{2}-\frac{1}{c^{2}}\frac{\partial^{2}}{\partial t^{2}}\right)
E_{probe}(\overrightarrow{r},t)=\frac{4\pi\rho}{c^{2}}\frac{\partial^{2}%
}{\partial t^{2}}\left\langle \widehat{d}\right\rangle _{probe}\text{.}
\tag{6}\label{6}%
\end{equation}
$\rho$ is the atom number density and $\left\langle \widehat{d}\right\rangle
_{probe}$ is the atomic dipole moment induced by a probe field. To acquire the
latter, one has to find first the atomic state $\left\vert \Psi
(z,t)\right\rangle $ in a combined field of pump(dressing) and probe fields,
then implement the ordinary quantum mechanical averaging of the dipole
operator by means of this state vector, and later select terms proportional to
the probe field $E_{probe}(\overrightarrow{r},t)$. The probe field is
$E_{probe}(\overrightarrow{r},t)=\epsilon_{0}(\overrightarrow{r}%
,t)\exp[i\overrightarrow{k}\overrightarrow{r}-i\omega t]+c.c.$ with slowly
varying amplitude $\epsilon_{0}(\overrightarrow{r},t)$.

The atomic state vector in combined (pump + probe) field has the form%
\begin{equation}
\left\vert \Psi(z,t)\right\rangle =\left\vert \Psi(z,t)\right\rangle
_{pump}+\left\vert \Delta\Psi(z,t)\right\rangle =(f(t)+C_{1}(z,t))\left\vert
1\right\rangle e^{-\frac{i}{\hbar}E_{1}t}+(g(t)+C_{2}(z,t))\left\vert
2\right\rangle e^{-\frac{i}{\hbar}E_{2}t}\text{,} \tag{7}\label{7}%
\end{equation}
where $f(t-z/c)$ and $g(t-z/c)$ are the above determined probability
amplitudes in pump laser field. \ The additional terms $C_{1}(\overrightarrow
{r},t)$ and $C_{2}(\overrightarrow{r},t)$ arise due to interaction with probe
radiation and are proportional to the probe field intensity in frame of linear
theory. \ Note that in distinction from [14], where the pump field was assumed
to be strictly monochromatic, here the problem of atomic states in the field
of pump radiation is time dependent.

\ Also it should be noted here that in the asymtotic case when $t-z/c>>T$%
\[
f(t-z/c)\approx\exp\left[  \left(  \frac{t-z/c}{T}\right)  \left(
-iV_{0}T-a\right)  \right]  h_{1}+\exp\left[  \left(  \frac{t-z/c}{T}\right)
\left(  -iV_{0}T-b\right)  h_{2}\right]
\]%
\[
g(t-z/c)\approx e^{-i\Delta(t-z/c)}\exp\left[  \left(  \frac{t-z/c}{T}\right)
\left(  -iV_{0}T-a\right)  \right]  (h_{1}-h_{3})+\exp\left[  \left(
\frac{t-z/c}{T}\right)  \left(  -iV_{0}T-b\right)  h_{2}\right]  (h_{2}%
-h_{4})
\]
with $h_{1}=\Gamma(c)\Gamma(b-a)/\Gamma(b)\Gamma(c-a)$, $h_{2}=\Gamma
(c)\Gamma(a-b)/\Gamma(a)\Gamma(c-b)$, $h_{3}=\Gamma(c+1)\Gamma(b-a)/\Gamma
(b+1)\Gamma(c-a)$, $h_{4}=\Gamma(c+1)\Gamma(a-b)/\Gamma(a+1)\Gamma(c-b\dot{)}%
$. So the switching on of the field brings to formation of 4 adiabatic terms,
which have different energies in distinction to the case discussed in [14]
where were assume only 2 terms with different energies.

The Hamiltonian in dipole approximation is:%
\begin{equation}
\widehat{H}=\widehat{H}_{0}-\widehat{\overrightarrow{d}}\overrightarrow{\text{
}E}_{pump}-\widehat{\overrightarrow{d}}\overrightarrow{\text{ }E}_{probe}
\tag{8}\label{8}%
\end{equation}

After standard calculations in frame of Schr\H{o}dinger equation, for
$C_{1}(\overrightarrow{r},t)$ and $C_{2}(\overrightarrow{r},t)$ amplitudes we
obtain%
\begin{equation}
C_{1}(\overrightarrow{r},t)=\frac{i}{\hbar}\overrightarrow{d}_{12}\left(
e^{i\overrightarrow{k}\overrightarrow{r}}\sigma_{1}(t)\overrightarrow
{\varepsilon}_{0}(t)+e^{-i\overrightarrow{k}\overrightarrow{r}}\sigma
_{2}(t)\overrightarrow{\varepsilon}_{0}^{\ast}(t)\right)  \tag{9}\label{9}%
\end{equation}
and%
\begin{equation}
C_{2}(\overrightarrow{r},t)=\frac{i}{\hbar}\overrightarrow{d}_{12}^{\ast
}\left(  e^{i\overrightarrow{k}\overrightarrow{r}}\theta_{2}(t)\overrightarrow
{\varepsilon}_{0}(t)+e_{2}^{-i\overrightarrow{k}\overrightarrow{r}}\theta
_{1}(t)\overrightarrow{\varepsilon}_{0}^{\ast}(t)\right)  \text{,}
\tag{10}\label{10}%
\end{equation}
where%
\[
\left\{  \sigma_{1},\theta_{1}\right\}  =%
%TCIMACRO{\dint \limits_{t_{0}}^{t}}%
%BeginExpansion
{\displaystyle\int\limits_{t_{0}}^{t}}
%EndExpansion
\left\{  \beta,\alpha\right\}  \text{ }e^{\mp i(\omega+\omega_{0})t^{\prime}%
}dt^{\prime}\text{,}%
\]%
\[
\{\sigma_{2},\theta_{2}\}=%
%TCIMACRO{\dint \limits_{t_{0}}^{t}}%
%BeginExpansion
{\displaystyle\int\limits_{t_{0}}^{t}}
%EndExpansion
\left\{  \beta,\alpha\right\}  \text{ }e^{\mp i(\omega-\omega_{0})t^{\prime}%
}dt^{\prime}\text{,}%
\]
Insertion of found state vectors (7) into $\left\langle \widehat
{d}\right\rangle _{probe}=$ $\left\langle \Psi(z,t)\right\vert \widehat{d}%
$\ $\left\vert \Psi(z,t)\right\rangle $ determines the right-hand side of wave
Eq.(6) as an explicit function of system parameters, proportional to the probe
wave amplitude.

In the next step we apply the well known slowly varying approximation to the
left-hand side of equation (6) and thus arrive to its reduced form, which is a
first order differential equation for the probe wave amplitude $\epsilon
_{0}(\overrightarrow{r},t)$ with partial derivatives in both, space and time
variables. \ Some of the right-hand side terms of obtained reduced wave
equation (rwe) are responsible for hyper-Raman scattering, parametric down
conversion and four-wave parametric amplification, respectively. However in
this paper these processes will not be considered and we will focus our
attention on the main nonparametric propagation process. In familiar
formulation this approach, as is well known, leads to determination of the
medium refractive index.

Introducing new variables \ $\tau=t-(\overrightarrow{k}\overrightarrow
{r})/\omega$ and $\eta=$ $(\overrightarrow{k}\overrightarrow{r})/k$, we
transform our reduced wave equation into an ordinary equation relative to
$\eta$ variable where $\tau$ appears as a parameter and thus rwe can be easily
integrated. \ Assuming that the incident probe wave repeats the form of the
pump one and propagates along the pump direction ($\eta=z$) we arrive to the
following simple expression for the seeking probe field amplitude
$\epsilon_{0}(z,\tau)$:%
\begin{equation}
\epsilon_{0}(z,\tau)=\frac{\epsilon_{0}}{\left(  1+e^{-\tau/T}\right)  }%
\exp\left[  \frac{2\pi\rho\omega_{0}^{2}}{\hbar c\omega}\left\vert
d_{12}\right\vert ^{2}%
%TCIMACRO{\dint \limits_{0}^{z}}%
%BeginExpansion
{\displaystyle\int\limits_{0}^{z}}
%EndExpansion
\left(  f^{\ast}(\tau)\theta_{1}(\widetilde{z},\tau)-g(\tau)\sigma
_{1}(\widetilde{z},\tau)\right)  e^{i(\omega-\omega_{0})(\tau+\widetilde
{z}/c)}d\widetilde{z}\right]  \text{.} \tag{11}\label{11}%
\end{equation}
Expression of $\epsilon_{0}(z,\tau)$ is the main product of this paper. It
concretizes the result of [14] in case of time dependent pumping field
creating the necessary for QS generator \ quantum superposition of ground and
excited states from the initial state. The imaginary part of \ (11) stipulates
a phase modulation, while the real part introduces amplitude modulation and
intensity variance of the probe laser beam during the propagation in the
medium. On the other hand, the exponent is a periodic function of time and
spatial coordinate, which results in a periodic-type modulation of the probe
field intensity, accompanied by amplification or weakening in average.

\section{Results and Conclusion}

To conceive roughly the picture of probe wave modulation developing by (11)
lets turn to the two-level model of alkali metal gases. The characteristic
values of $\left\vert d\right\vert ^{2}$ for dipole allowed transitions are
around $2\times10^{-34}CGSE$ and sample concentration can be varied in a wide
range of $10^{12}-10^{16}cm^{-3}$. Typical line broadening is $10^{7}%
-10^{8}Hz$ and therefore the lowest allowed in frame of this model value for
resonance detuning is $\Delta=2\ast10^{8}Hz$. A picture of probe wave
modulation under some possible conditions is given in Fig. 1. In particular
here we have presented the case when $t>>T$, that is when the incident field
has stabilized. It shows the ability of the \ QS generator \ scheme to
juxtapose the composition of high repetition ultrashort pulses with essential
\ amplification in the frame of chosen manner of state preparation.

A close consideration of the exponential in Eq. (11) shows that the
regularities of QS generator are very simple and convenient from
experimental/applied viewpoint. \ The rate of probe wave modulation, for
instance, is determined solely by the generalized Rabi frequency $\Omega
=\sqrt{\Delta^{2}+4V_{0}^{2}}$. This rate in fact determines the space and
time repetition distance \ between the pulses: $\Delta t_{repetition}%
=2\pi/\Omega$ and $\Delta z_{repetition}=c\cdot$ $\Delta t_{repetition}$.
\ The regime of propagation, amplification or weakening, is determined by the
detuning $\omega-$ $\omega_{pump}$ (See Fig.1 in [14]). \ This dependence is
especially sharp near the scattering resonances. The product of gas density on
a single-photon scattering cross section determines the modulation depth.
Thus, when in amplification regime, increase of the gas density deepens the
modulation and thus results in narrowing of peaks in the train.

To incorporate the damping phenomena into theory we will use a simple method,
that is we will add to the transition frequency $\omega_{0}$ in (11) the
complex quantity $i\gamma$ where $\gamma$ for discusssed parameters is of the
order of $10^{7}Hz$. As long as we have a resonance detuning greater than the
line broadening, this procedure describes the damping phenomena very well. Our
approach takes into account also the spontaneous damping of excitation. \ \ 

Another factor which should be taken into account is the Doppler or
inhomogeneous broadening of optical transition. In a dilute gas at room or
higher temperatures the Doppler linewidth prevails the natural and collisional
linewidths. We assume a Maxwell-Boltzmann velocity distribution in laser
propagation direction. \ To actually calculate the influence \ of Doppler
broadening we should add $\Delta_{Doppler}$ to the detuning and then average
over the velocity distribution [17]. The results of these calculations carried
out for the same conditions as in Fig.1, and including the relaxation process
and the Doppler broadening, prove our assertion that relaxations have minor
role in QS generator when far from homogeneous broadening of spectral lines
and that the Doppler broadening can not destroy the pulse train formation
under appropriate conditions.%

%TCIMACRO{\FRAME{ftbpFU}{3.8222in}{3.1072in}{0pt}{\Qcb{Modulation and
%amplification of the probe wave in a two-level media. \ Here $\Delta
%=-5\times10^{12}Hz$, $\omega_{0}=10^{15}Hz$, $\omega-\omega_{0}=5.004\times
%10^{11}Hz$,$\rho=10^{15}cm^{-3}$ $z=i\pi c/(b-a)$ and switching characteristic
%duration $T$ is $10^{-12}s.$}}{\Qlb{Fig1}}{figure1.eps}%
%width 3.8222in;  height 3.1072in;  depth 0pt;  original-width 3.8152in;
%original-height 3.0968in;  cropleft "0";  croptop "1";  cropright "1";
%cropbottom "0";  filename 'Figure1.eps';file-properties "XNPEU";}} }%
%BeginExpansion
\begin{figure}
[ptb]
\begin{center}
\includegraphics[
height=3.1072in,
width=3.8222in
]%
{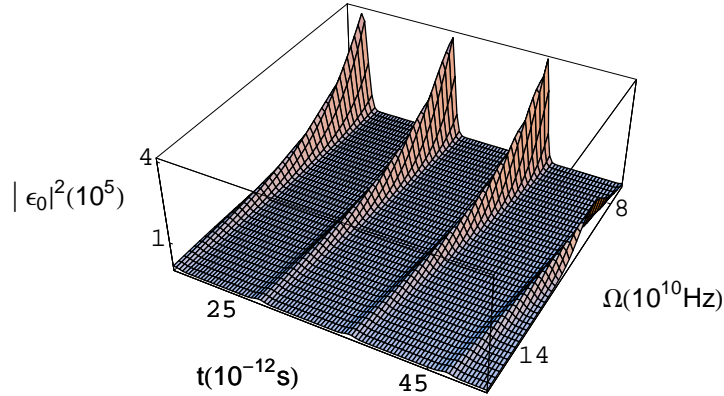}%
\caption{Modulation and amplification of the probe wave in a two-level media.
\ Here $\Delta=-5\times10^{12}Hz$, $\omega_{0}=10^{15}Hz$, $\omega-\omega
_{0}=5.004\times10^{11}Hz$,$\rho=10^{15}cm^{-3}$ $z=i\pi c/(b-a)$ and
switching characteristic duration $T$ is $10^{-12}s.$}%
\label{Fig1}%
\end{center}
\end{figure}
%EndExpansion

In conclusion, we have shown that \ the rapid switching-on of the pump field
intensity in a two-level atomic medium may ensure a mixing of adiabatic terms
in a way sufficient for formation of the QS generator. The repetition rate and
duration of pulses are easily regulated by means of smooth changing of the
pump field resonance detuning \ and atomic concentration respectively.
\ Numerical calculations \ (for alkali metal vapors) show that the presented
mechanism of QS generator of ultrashort pulses is robust against \ the
homogeneous and inhomogeneous broadening of spectral lines in a very wide
range of parameters, as well as parameter fluctuations.

\begin{acknowledgement}
Authors thank Atom Zh. Muradyan for helpful discussions. \ This work was
supported by Alexander von Humboldt foundation and Armenian Science Ministry
Grant 143.
\end{acknowledgement}

\end{document}